\documentclass[useAMS,usenatbib]{mn2e}

\usepackage{graphicx,natbib,epsfig,psfig}
\usepackage{amsmath}
\usepackage{amssymb}

\def\bc{\begin{center}}
\def\ec{\end{center}}

\title{Merging of Low-Mass Systems and~ \\ 
the Origin of the Fundamental Plane}

\author[E.A.~Evstigneeva et al.]
{E.A.~Evstigneeva$^{1,2}$, R.R.~de Carvalho$^{3}$, A.L.~Ribeiro$^4$, and 
H.V.~Capelato$^5$\\
$^1$Astronomical Institute of St.Petersburg State University,
198504 St.Petersburg, Russia\\
$^2$Present address: Department of Physics, University of Queensland, QLD 4072,
Australia, katya@physics.uq.edu.au\\
$^3$Observat\'orio Nacional, 20921-400 Rio de Janeiro, Brazil\\
$^4$Universidade Estadual de Santa Cruz, 45650-000, Ilheus-BA, Brazil\\
$^5$Divis\~ao de Astrof\'{\i}sica - INPE/MCT CP 515, 12201-970 SP, Brazil}

\pagerange{000-000} \pubyear{2004}

\begin{document}

\maketitle

\begin{abstract}
We present a new set of dissipationless N-body simulations to
examine the feasibility of creating bright ellipticals (following the Kormendy
relation) by hierarchically merging present-day early-type dwarf galaxies,
and to study how the encounter parameters affect the location
of the end-product in the $\langle \mu_{e} \rangle - R_{e}$ plane.  
We investigate the merging of one-component galaxies of both equal and 
different masses, the merging of two-component galaxy models to explore the 
effect of dark halos on the final galaxy characteristics, and the merging of 
ultracompact dwarf galaxies. We find that the
increase of $\langle \mu_{e} \rangle$ with $R_e$ is attributable to an increase 
in the initial orbital energy.  The merger remnants shift down in the 
$\langle \mu_{e} \rangle - R_{e}$ plane and fail to reach the KR. Thus, the KR 
is not reproducable by mergers of dwarf early-type systems, rendering untenable 
the theory that present-day dwarfs are responsible for even a small fraction of 
the present-day ellipticals, unless a considerable amount of dissipation is 
invoked. However, we do find that present-day dwarfs can be formed by the 
merger of ultra-compact dwarfs.

\end{abstract}


\section{Introduction}

Despite the progress made in understanding the physics of early-type galaxies, a single
widely accepted formation model, capable of explaining all their observed properties, does
not yet exist. There is a variety of scenarios describing how these objects were formed,
ranging from the collapse of clumpy protogalaxies (Eggen, Lynden-Bell \& Sandage
1962, Larson 1975) to the merging of smaller galaxies (White \& Rees
1978, Kauffmann 1996, Cole et al. 2000).  Nevertheless, the
remarkable fact is that early-type galaxies demonstrate a very tight
kinematic-structural relationship, usually referred to as the Fundamental Plane 
(Djorgovski \& Davis 1987, Dressler et al. 1987), and any theory of galaxy formation
and evolution must be able to account for its tightness and environmental independence
(de la Rosa, de Carvalho, \& Zepf 2001).

The Fundamental Plane (FP) combines photometric parameters ($R_e$, the
effective radius, along with 
$\langle \mu_{e} \rangle$, the mean surface brightness within $R_e$)
with a spectroscopic observable (the line-of-sight central velocity
dispersion $\sigma_0$). The measured values of $R_e$, $\mu_e$ and {$\sigma_0$ 
for a sample of E and S0 galaxies fill a thin plane (with
scatter of $\sim$ 0.1 dex) within this 3-parameter space instead of
spreading out over the whole space.  The FP can be projected onto any
pair of axes from the three variables. Examples of these
projections are the Kormendy relation: $\mu_e$--log$R_e$ (Kormendy
1977) and the Faber-Jackson relation between luminosity
and central velocity dispersion (Faber \& Jackson 1976).

\begin{figure*}
\centerline{\psfig{file=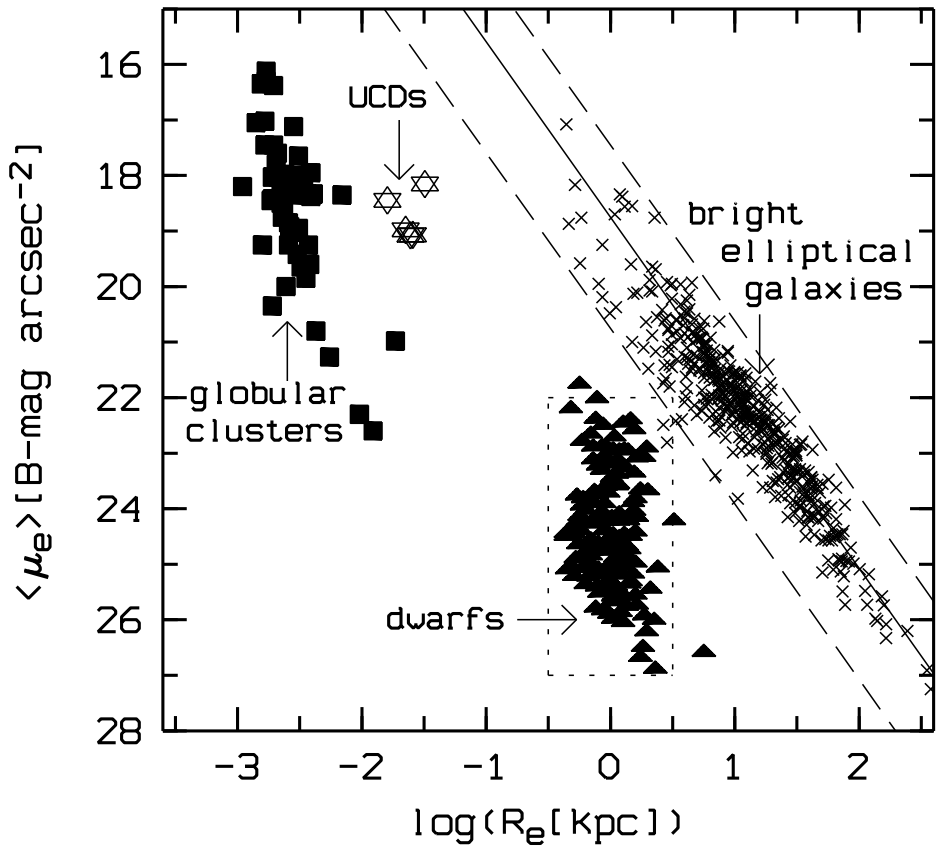,angle=-90,width=14cm,clip=}} 
\caption{The $\langle
\mu_{e} \rangle - R_{e}$ plane for different types of objects. Filled triangles and
crosses are data from Capaccioli, Caon \& D'Onofrio (1992). Filled squares
are globular clusters from Burstein et al. (1997). Starred symbols are
ultracompact dwarf galaxies from Drinkwater et al. (2003). The solid line
represents the Kormendy relation. Both dashed lines represent the locii of families
exhibited in this diagram.  They schematically represent their overall extension.}
\end{figure*}

The distribution of elliptical galaxies in the $\langle \mu_{e} \rangle - R_{e}$ plane,
shown in Figure 1, reveals the existence of two distinct families: the ``ordinary''
family, composed of early-type dwarfs, lying in the region defined by $22 \leq \langle
\mu_e(B) \rangle \leq 27$ and $R_e \leq 3$ kpc and which are always fainter than $M(B)
= -19.3$, and the ``bright'' family, containing the most luminous elliptical galaxies
(the brightest cluster galaxies - BCGs, which host QSOs and Seyfert nuclei),
populating a relatively narrow strip in the $\langle \mu_{e} \rangle - R_{e}$ plane,
with $M(B) < -19.3$ and extending over 2 dex in $R_e$.  The presence of boxy
isophotes in these luminous galaxies and the high frequency of multiple nuclei among
BCGs lends support to the merging scenario for the ``bright'' family.  
The low-luminosity disky ellipticals occupy the region between dwarfs and 
bright elliticals in the $\langle \mu_{e} \rangle - R_{e}$ plane 
($-0.5 \lesssim \rm{log}R_e \lesssim 0.7$,  $18 \lesssim \langle \mu_e(B) \rangle \lesssim 23$).
We did not show them in the figure to avoid confusion.
Capaccioli, Caon
\& D'Onofrio (1992) interpreted the $\langle \mu_{e} \rangle - R_{e}$
plane for ellipticals as a logical equivalent to the HR diagram for stars, and
that galaxies might reach the ``bright'' end of this diagram through successive mergers
of low-mass systems.

The segregation of galaxies in the $\langle \mu_e \rangle - R_e$ plane, together with
the different behavior of dEs and luminous elliptical galaxies (Es) in the $M(B) -
\langle \mu_e \rangle$ diagram, where $M(B)$ is the absolute
magnitude, (see Graham \& Guzman 2003), was generally interpreted as evidence
for distinct formation and/or evolutionary processes for the two families of objects.
However, there is evidence for a continuity, rather than dichotomy, between Es and dEs. 
According to Graham \& Guzman (2003), dE galaxies form a
continuous sequence with the brighter E galaxies, such that $\mu_0$ (the central surface
brightness) brightens linearly with $M(B)$ until core formation causes the most
luminous E galaxies ($M(B) \leq -20.5$) to deviate from this relation. The different
behavior of dE and E galaxies in the $M(B) - \langle \mu_e \rangle$ (or $M(B) - \mu_e$)
diagram, and the $\langle \mu_e \rangle - R_e$ plane, is unrelated to core
formation, and is expected from the continuous and linear relation between $M(B)$ and
$\mu_0$, and $M(B)$ and log(n) (the Sersic index).

In hierarchical galaxy formation scenarios (as in the CDM model), small
galaxies (or dark matter halos) are the building blocks of more
massive galaxies, and should have been more numerous in the early
Universe. Thus, the present-day dwarfs can be seen as survivors of an
initially much richer population. However, we observe many fewer dwarfs
today than the predicted number of surviving dark matter halos (Klypin
et al. 1999, Moore et al. 1999).  There also appear
to be inconsistencies in the timescales necessary to construct large
galaxies (Prantzos \& Silk 1998), and differences in the
stellar populations of large and small galaxies (Tolstoy et
al. 2003).  These problems point to a need to better
understand the physics of dwarf galaxies.

\begin{table*}
\caption[1]{The initial simulation parameters and the relative variations of the total
energy and the total angular momentum} \bc
\begin{tabular}{lllllll}
\hline
\\
 & $\epsilon$ & $\Delta t$ &  $N$ & $M$ & $\Delta E/E \, \%$ &
 $\Delta J/J \, \%$ \\ \\
\hline
\\
Low-mass systems: & & & & & & \\ \\
One-component & $\epsilon_{1,2} = 0.07$& 0.025 &
$N_{1,2} = 90\,000$ & $M_{1,2} = 9$ & 0.4 & 2.8 \\
equal mass merging & & & & & & \\ \\
One-component & $\epsilon_1 \,\,\,\,= 0.09$ & 0.1 & $N_1 \,\,\,\,= 90\,000$ &
$M_1 \,\,\,\,= 9$ & 0.7 & 2.3 \\
different mass merging & $\epsilon_2 \,\,\,\, = 0.13$ &  &
$N_2 \,\,\,\,= 30\,000$ & $M_2 \,\,\,\,= 3$ & & \\ \\
UCDs & $\epsilon_{1,2} = 0.002$& 0.001 &
$N_{1,2} = 30\,000$ & $M_{1,2} = 0.6$ & 5.0 & 2.0 \\ \\
Two-component & $\epsilon_s \,\,\,\,= 0.15$ & 0.15 &
$N_s \,\,\,\,= 20\,000$ & $M_s \,\,\,\,= 9$ & 2.5 & 1.5 \\
equal mass merging & $\epsilon_h \,\,\,\,= 0.40$ &  & $N_h \,\,\,\,= 60\,000$ &
$M_h \,\,\,\,= 27$ & & \\ \\
\hline \\
Merging of ellipticals & $\epsilon_{1,2} = 0.07$ & 0.005 &
$N_{1,2} = 90\,000$ & $M_{1,2} = 2\,229$ & 3.0 & 1.8 \\
lying on the KR & & & & & & \\ \\
Merging of objects & $\epsilon_{1,2} = 0.07$ & 0.001 &
$N_{1,2} = 180\,000$ & $M_{1,2} = 35\,218$ & 1.3 & 4.0 \\
lying above the KR & & & & & & \\ \\
\hline
\end{tabular}
\ec
\end{table*}

In the present paper, we concentrate on a single aspect of the complex properties
of this family of galaxies, the dwarf ellipticals found in the lower left
corner of Figure 1 ($\langle \mu_e(B) \rangle \geq 22$ and $R_e \leq 3$ kpc). Dwarf
ellipticals (dEs) are spherical or elliptical in appearance, compact, with high
central stellar densities.  They are fainter than $M(B) = -18$ and have low masses
($M_{tot} \leq 10^9~M_{\odot}$).  They are found preferentially in the vicinity of massive galaxies,
usually have little or no detectable gas, and are often not rotationally supported.
When gas is detected, it exhibits an asymmetric distribution, is less extended than the
underlying stellar component, and seems to be kinematically distinct (see
Grebel 2001 and references therein).  dEs sometimes
show pronounced nuclei, with the fraction of nucleated dEs increasing with luminosity. 
The surface density profiles of dEs are best described by Sersic's generalization
of de Vaucouleurs $R^{1/4}$ law and exponential profiles (see Grebel 2001 and
references therein).  The origin of these early-type dwarfs is still unknown.  Three general
classes of models have been suggested (Grebel 2003): tidal interactions that
transform field disk galaxies into spheroidal systems; processes associated with the
birth of small systems (in the CDM model); and fragments torn from collisions between
larger galaxies. However, each of these models encounter certain difficulties in reproducing the observed
properties of dE's.

The effects of dissipationless merging on the FP have been explored in
Capelato, de Carvalho \& Carlberg (1995),
Dantas et al. (2003), and Nipoti, Londrillo \& Ciotti
(2003), among others. In these contributions, the progenitors lay on the FP, and
the edge-on projection of the FP was found to be reproduced by
dissipationless merging.  In contrast to these works, our initial
models are low-mass early-type galaxies which lie neither on the
FP nor along the KR.
We consider the merging of
one-component galaxies of both equal and different masses, as well as the merging of
two-component galaxy models, to explore the effects of dark halos on the
final galaxy characteristics. We also examine mergers of ultracompact
dwarf galaxies (UCDs), as these constitute a recently discovered class of very
compact objects (Drinkwater et al. 2001, Bekki et al.
2001), populating the FP in a previously empty region between the
most luminous globular clusters and nucleated dwarf galaxies.

To form a complete picture and compare our results with those obtained by other
authors, we run a series of simulations where the progenitor models lie on the KR. We also merge
objects which lie above the KR in the $\langle \mu_{e} \rangle - R_{e}$ plane.  These
objects are not found in the nearby universe. but may have existed at earlier
epochs, and are probable progenitors for present-day ellipticals.

\section{Numerical method and input}

\subsection{The numerical code}

Our computations utilize an N-body code with a hierarchical tree algorithm and
multipole expansion to compute the forces, as proposed by Barnes \& Hut
(1986). The force computation includes the quadrupole correction terms,
following Dubinski (1988).

The set of input parameters ($\epsilon$, the potential softening length; $N$, the number of particles; 
and $\Delta t$, the integration timestep)
used in the various simulations are listed in Table 1. Also given in
this Table are the upper limits of the relative variation of the total
energy, $E$, and the relative variation of the total angular momentum, $J$, for
simulations using each set of parameters. The tolerance parameter was the same for
all simulations, $\theta$ = 0.8. As discussed in Dantas et al. (2002), 
the choice of the
softening parameter $\epsilon$ is a compromise between spatial resolution and the
collisionless condition of the system on evolutionary time scales. It thus depends
on the particle number $N$, and on the specific density distribution profile (see
also e.g. Merritt 1996).

\begin{table*}
\caption[2]{Merging of one-component equal mass models} \bc
\begin{tabular}{llllllll}
\hline
\\
Run & $\hat E$ & $\hat L$ & $T_{per}$ & $20\,T_{cr}$ & $T_{end}$ & log$R_e$(kpc) & $\langle \mu_{e} \rangle$ \\ \\
\hline
\\
progenitor($dw_1$) & & & & & & 0.19+/-0.00 & 25.00+/-0.01 \\ \\
E01 & -3 & 0 &   4.2 & 144 & 275 & 0.43+/-0.02 & 25.48+/-0.11 \\
E02 & -7 & 0 &   4.3 & 100 & 100 & 0.34+/-0.02 & 25.00+/-0.10 \\
E03 & -5 & 0 &   5.3 & 119 & 121 & 0.38+/-0.02 & 25.23+/-0.12 \\
E04 & -3 & 1 &  24.3 & 140 & 150 & 0.43+/-0.02 & 25.50+/-0.11 \\
E05 & -5 & 1 &  11.3 & 117 & 120 & 0.38+/-0.03 & 25.21+/-0.13 \\
E06 & -7 & 1 &   6.8 & 102 & 120 & 0.34+/-0.02 & 25.02+/-0.10 \\
E07 & -3 & 2 &  24.3 & 140 & 150 & 0.43+/-0.02 & 25.46+/-0.11 \\
E08 & -5 & 2 &  11.6 &  91 & 150 & 0.38+/-0.02 & 25.20+/-0.10 \\
E09 & -7 & 2 &   6.8 & 102 & 175 & 0.34+/-0.02 & 25.00+/-0.10 \\
E10 & -1 & 3 & 126.3 & 174 & 200 & 0.47+/-0.02 & 25.70+/-0.12 \\
E11 & 0 & 1 & 2.6 & 166 & 170 & 0.49+/-0.03 & 25.87+/-0.15 \\
E12 & 0.5 & 1 & 2.5 & 178 & 182 & 0.49+/-0.03 & 25.87+/-0.15 \\
E13 & 1 & 1 & 2.5 & 191 & 200 & 0.50+/-0.03 & 25.95+/-0.14 \\
 \\
E01$_2$ & -3 & 0 &  5.8 & 193 & 234 & 0.62+/-0.03 & 26.76+/-0.16 \\
E02$_2$ & -7 & 0 &  4.2 &  82 & 150 & 0.47+/-0.05 & 24.93+/-0.24 \\
E05$_2$ & -5 & 1 & 11.6 & 109 & 150 & 0.54+/-0.04 & 25.27+/-0.20 \\
E06$_2$ & -7 & 1 &  7.0 &  83 & 125 & 0.46+/-0.04 & 24.87+/-0.18 \\
E08$_2$ & -5 & 2 & 11.6 &  96 & 231 & 0.55+/-0.04 & 25.33+/-0.20 \\
E09$_2$ & -7 & 2 &  7.0 &  83 & 150 & 0.47+/-0.04 & 24.90+/-0.19 \\
\\
E01$_3$ & -3 & 0 &  8.9 & 416 & 546 & 0.83+/-0.03 & 26.19+/-0.17 \\
E02$_3$ & -7 & 0 &  4.2&  85 & 150 & 0.59+/-0.05 & 24.78+/-0.26 \\
E05$_3$ & -5 & 1 & 11.6 &  84 & 200 & 0.71+/-0.04 & 25.37+/-0.22 \\
E06$_3$ & -7 & 1 &  7.0 & 134 & 359 & 0.58+/-0.05 & 24.71+/-0.24 \\
E09$_3$ & -7 & 2 &  7.0&  87 & 300 & 0.59+/-0.06 & 24.75+/-0.27 \\ \\
\hline
\end{tabular}
\ec
\end{table*}

\begin{table*}
\caption[3]{Merging of one-component different mass progenitors} \bc
\begin{tabular}{llllllll}
\hline
\\
Run & $\hat E$ & $\hat L$ & $T_{per}$ & $20\,T_{cr}$ & $T_{end}$ & log$R_e$(kpc) & $\langle \mu_{e} \rangle$ \\ \\
\hline
\\
progenitor1($dw_1$) & & & & & & 0.19+/-0.00 & 25.00+/-0.01 \\
progenitor2($dw_2$) & & & & & & 0.18+/-0.00 & 26.16+/-0.01 \\ \\
D01 & -3 & 1 & 35.0 & 110 & 164 & 0.36+/-0.02 & 25.60+/-0.10 \\
D02 & -5 & 1 & 16.3 &  99 & 152 & 0.34+/-0.02 & 25.46+/-0.09 \\
D03 & -7 & 1 &  9.9 &  91 & 137 & 0.30+/-0.01 & 25.28+/-0.07 \\
D04 & -3 & 2 & 35.0 & 110 & 164 & 0.36+/-0.02 & 25.58+/-0.10 \\
D05 & -5 & 2 & 16.3 & 100 & 152 & 0.33+/-0.02 & 25.44+/-0.09 \\
D06 & -7 & 2 &  9.8 &  92 & 140 & 0.30+/-0.02 & 25.28+/-0.08 \\ \\
\hline
\end{tabular}
\ec
\end{table*}

\begin{table*}
\caption[4]{Merging of ultracompact dwarf galaxies} \bc
\begin{tabular}{llllllll}
\hline
\\
Run & $\hat E$ & $\hat L$ & $T_{per}$ & $20\,T_{cr}$ & $T_{end}$ & log$R_e$(kpc) & $\langle \mu_{e} \rangle$ \\ \\
\hline
\\
progenitor($ucd$) & & & & & & -1.65+/-0.00 & 18.75+/-0.01 \\ \\
F01 & -1 & 1 &     0.84 & 0.96 & 2.0 & -1.32+/-0.03 & 19.69+/-0.13 \\
F01$_2$ & -1 & 1 & 1.80 & 1.80 & 4.0 & -1.03+/-0.04 & 20.44+/-0.21 \\
F01$_3$ & -1 & 1 & 3.22 & 3.55 & 7.3 & -0.74+/-0.05 & 21.14+/-0.23 \\
F01$_4$ & -1 & 1 & 5.78 & 6.82 & 18.5 & -0.47+/-0.04 & 21.80+/-0.22 \\ \\
\hline
\end{tabular}
\ec
\end{table*}

\begin{table*}
\caption[5]{Merging of two equal mass progenitors with dark halo} \bc
\begin{tabular}{llllllll}
\hline
\\
Run & $\hat E$ & $\hat L$ & $T_{per}$ & $20\,T_{cr}$ & $T_{end}$ & log$R_e$(kpc) & $\langle \mu_{e} \rangle$ \\ \\
\hline
\\
progenitor & & & & & & 0.19+/-0.00 & 25.01+/-0.01 \\ \\
H01 & -3 & 1 & 55.8 & 245 & 700 & 0.37+/-0.03 & 25.34+/-0.17 \\
H02 & -5 & 1 & 26.0 & 204 & 700 & 0.36+/-0.04 & 25.31+/-0.17 \\
H03 & -7 & 1 & 15.7 & 174 & 600 & 0.35+/-0.04 & 25.26+/-0.18 \\ \\
\hline
\end{tabular}
\ec
\end{table*}

\begin{table*}
\caption[6]{Merging of ellipticals lying on the KR} \bc
\begin{tabular}{lllllll}
\hline
\\
Run & $\hat E$ & $\hat L$ &  $20\,T_{cr}$ & $T_{end}$ & log$R_e$(kpc) & $\langle \mu_{e} \rangle$ \\ \\
\hline
\\
progenitor($KR$) & & & & & 0.18+/-0.00 & 19.00+/-0.01 \\ \\
P01 & -3 & 1 & 7 & 18 & 0.44+/-0.03 & 19.54+/-0.13 \\
P02 & -5 & 1 & 6 & 18 & 0.39+/-0.02 & 19.30+/-0.11 \\
P03 & -7 & 1 & 5 & 18 & 0.35+/-0.02 & 19.08+/-0.11 \\
\\
P01$_2$ & -3 & 1 & 11 & 22 & 0.66+/-0.04 & 19.92+/-0.20\\
P02$_2$ & -5 & 1 & 8 & 24 & 0.56+/-0.04 & 19.41+/-0.19 \\
P03$_2$ & -7 & 1 & 6 & 20 & 0.49+/-0.04 & 19.01+/-0.21 \\ \\
\hline
\end{tabular}
\ec
\end{table*}

\begin{table*}
\caption[7]{Merging of ellipticals lying above the KR in the $\langle \mu_{e} \rangle -
R_{e}$ plane} \bc
\begin{tabular}{lllllll}
\hline
\\
Run & $\hat E$ & $\hat L$ &  $20\,T_{cr}$ & $T_{end}$ & log$R_e$(kpc) & $\langle \mu_{e} \rangle$ \\ \\
\hline
\\
progenitor($AKR$) & & & & & 0.19+/-0.00 & 16.03+/-0.00 \\ \\
A04 & -3 & 1 & 1.74 & 8 & 0.44+/-0.02 & 16.55+/-0.12 \\
A05 & -5 & 1 & 1.46 & 6 & 0.39+/-0.02 & 16.28+/-0.11 \\
A06 & -7 & 1 & 1.42 & 6 & 0.34+/-0.02 & 16.05+/-0.10 \\
A07 & -3 & 2 & 1.74 & 8 & 0.44+/-0.02 & 16.53+/-0.11 \\
A08 & -5 & 2 & 1.46 & 6 & 0.38+/-0.02 & 16.26+/-0.11 \\
A09 & -7 & 2 & 1.26 & 6 & 0.34+/-0.02 & 16.04+/-0.11 \\ \\
\hline
\end{tabular}
\ec
\end{table*}

\subsection{Galaxy models}

We measure mass and length in units of $10^8~M_{\odot}$ and 1
kpc, respectively. These values, together with the gravitational constant $G = 1$, fix our
time and velocity units as 47.2 Myr and 20.7 km\,s$^{-1}$, respectively.

The initial models for merging of low-mass galaxies were chosen from the region in
Figure 1 occupied by ordinary ellipticals and early-type dwarfs (the
``ordinary'' group). Both one- and two-component models were constructed as described below.

The initial one-component galaxies were modelled using the potential-density pair for spherical galaxies
given in Hernquist (1990) :

\begin{align}
\Phi(r) & = - \frac{G M}{r + a} \, ,   \\
\rho(r) & = \frac{M}{2 \pi a^3} \, \frac{a^4}{r\,(r+a)^3}\, ,
\end{align}
where $M$ is the galaxy mass and $a$ is the scale length.

In the case of two-component models (the luminous galaxy and its dark matter halo), the
halo density profile was represented as a truncated isothermal sphere (Hernquist
1993):

\begin{equation}
\rho(r) = \frac{M_h}{2 \pi^{3/2}} \, \frac{\alpha}{r_c} \,
\frac{exp(-r^2/{r_c}^2)}{r^2+\gamma^2} \,
\end{equation}
Here $M_{h}$ is the halo mass, $\gamma$ is the core radius,
and $r_{c}$ is the cutoff radius. The normalization constant $\alpha$ is defined by

\begin{equation}
\alpha = (1 - \sqrt{\pi}q\,exp(q^2)[1 - erf(q)])^{-1} \, ,
\end{equation}
where $q = \gamma/r_c$. The cumulative mass profile, $M_{h}(r)$, and potential,
$\Phi(r)$, corresponding to $\rho(r)$ are

\begin{align}
M_h(r) & = \frac{2 M_h \alpha}{\sqrt{\pi}} \, \int\limits_{0}^{r/r_c} \,
\frac{x^2 exp(-x^2)}{x^2 + q^2}\,dx\, , \\
\Phi_h(r)& = -\frac{G M_h(r)}{r}  + \frac{G M_h(r) \alpha}{\sqrt{\pi}
r_c}\,Ei[-(r/r_c)^2 - q^2]\, ,
\end{align}
where $Ei(z)$ is an exponential integral.

We constructed single-component models representing early-type dwarfs galaxies with
total masses of $M_{dw_1} = 9$ and $M_{dw_2} = 3$, both having the
same scale length $a = 1$ ($R_e \approx 1.8\,a$). 

The two-component model for
early-type dwarfs consists of a luminous component 
identical to that for one-component $dw_1$ model described above, with the dark matter halo
 more massive ($M_{h}/M_{s} = 3$) and less concentrated ($\gamma/a = 3$) than
the stellar component. The dark component parameters for these models are $M_{h} = 27$,
$\gamma = 3$, and $r_{c} = 12$. Given the lack of observational data on dark matter
in dwarf galaxies, these models are consistent with current ideas on the
structure of these objects. 

The effective radii of UCDs range from 0.015 kpc (log$R_e = -1.82$) to 0.03 kpc
(log$R_e = -1.52$) and their mean surface brightness within $R_e$ ranges from
18 to 19.
The UCD radial profiles are well described by de Vaucouleurs profiles (Drinkwater et al.
2003). For the initial UCD models, we used a Hernquist sphere with $M_{ucd} = 0.6$ 
and $a_{ucd} = 0.014$.

Finally, for the progenitor models lying  on and above the KR, we constructed models 
with $M_{KR} = 2229$, and $M_{AKR} = 35218$, respectively, with both models 
having $a = 1$.

\subsection{Initial conditions}

The encounter of two non-rotating
spherical galaxies may be characterized by the dimensionless energy and angular
momentum:

\begin{align}
\hat E &= \frac{E_{orb}}{\frac{1}{2}\mu \overline{\langle v^2 \rangle}} \, ,
\\
\hat L &= \frac{L_{orb}}{\mu \overline{r_h}\, \overline{\langle v^2
\rangle}^{\frac{1}{2}}} \, ,
\end{align}
where $\overline{\langle v^2 \rangle} = \sqrt{\langle v_1^2 \rangle \langle v_2^2
\rangle}$, $r_h = \sqrt{r_{h 1} r_{h 2}}$. Here $\langle v^2 \rangle$ is the internal
mean-square velocity, $r_h$ is the half-mass radius of a galaxy (Binney \& Tremaine \cite{binney}). The 
indices denote
each of the initial galaxies and $\mu$ is the reduced mass of the system.

The time required for two galaxies to merge is a function of the initial position of
the binary orbit in the $(\hat E,\hat L)$ plane defined by equations (7) and (8)
(Binney \& Tremaine 1987). We considered only those pairs of $(\hat E,\hat L)$
which correspond to rapid mergers
(less than a Hubble time).

To completely determine the orbital elements of the encounter, we must define a third parameter:

\begin{equation}
A = \frac{2\,G M}{\overline{r_h}\overline{\langle v^2 \rangle}} \, ,
\end{equation}
which depends only on the dynamical structure of the initial galaxies. The initial separations and velocities for the encounters are selected from a grid of $\hat E,\hat
L, A$ values. The initial separation of the models was taken at the apocenter position
for bound orbits and $\sim 4r_h$ for unbound ones.

We follow the dynamical evolution of each merger until the resulting system is virialized, which typically occurs on a time scale shorter than 20
$T_{cr}$ after the first encounter between two galaxies, where $T_{cr}$ is the crossing time of the final object ($T_{cr} = G M^{5/2}/(2
E)^{3/2}$). The values for this timescale are given in
Tables 2-7, column (5).

Tables 2-7 summarize the initial conditions and characteristic parameters of the
simulations. In Table 2, the mergers of one-component equal mass galaxies are labelled as follows: 
\begin{itemize}
\item E01 - E13: \textit{first-generation mergers}: simulations of the
encounter of two single-component equal mass progenitors with different pairs of $(\hat
E,\hat L)$;
\item E01$_2$ - E09$_2$: \textit{second-generation mergers}: simulations of the
encounter of two identical end-products of first generation mergers from the previous
set of simulations. Progenitors are selected such that E0$i_2$ = (E0$i$ + E0$i$), with
$i$ = \{1,2,5,6,8,9\}~;
\item E01$_3$ - E09$_3$: \textit{third-generation mergers}:
encounters between two identical second generation mergers with the progenitors
selected such that ~E0$i_3$ = E0$i_2$ + E0$i_2$, with $i$ = \{1,2,5,6,9\}.
\end{itemize}

The other simulations (Tables 3-7) with similar subscripts but a different starting 
letter, are as follows:
\begin{itemize}
\item D - mergers of different mass one-component mass galaxies; 
\item F - mergers of UCD models; 
\item H - mergers of two-component equal mass galaxies; 
\item P - mergers of one-component models lying \textit{on} the KR; 
\item A - mergers of one component models lying  \textit{above} the KR. 
\end{itemize}

In Table 6, P01$_2$ -
P03$_2$ are the second generation mergers which, as for the E mergers, had
progenitors selected such that P0$i_2$ = (P0$i$ + P0$i$), with $i$ = \{1,2,3\}.
Similarly, in Table 4, F01$_j$ = (F01$_{j-1}$ + F01$_{j-1}$), where $j$ = \{2,3,4\}. 

Columns (2) and (3) in Tables 2-7 give the initial dimensionless orbital energy and
corresponding angular momentum. Columns (4) and (6) give $T_{per}$, the predicted
time interval before the two-body pericenter, and $T_{end}$, the total time elapsed
to the end of the simulation, respectively.

\begin{figure}
\centerline{\psfig{file=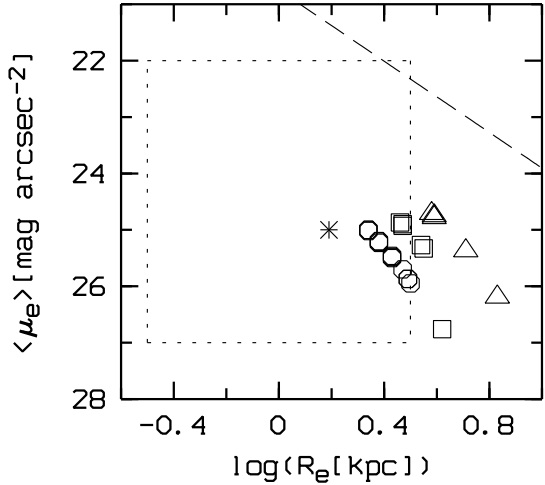,angle=-90,width=8cm,clip=}}
\caption{Merging of equal mass progenitors. * - the initial progenitor. Circles - the
1st generation progenitors; squares - the  2nd generation progenitors; triangles - the
3rd generation progenitors. The dashed line corresponds to the line in Fig.1 which
encompasses the region of bright ellipticals below and it is parallel to the KR. The
rectangular area embraces the locus of the dwarf galaxies (see Fig.1).}
\end{figure}

\begin{figure}
\centerline{\psfig{file=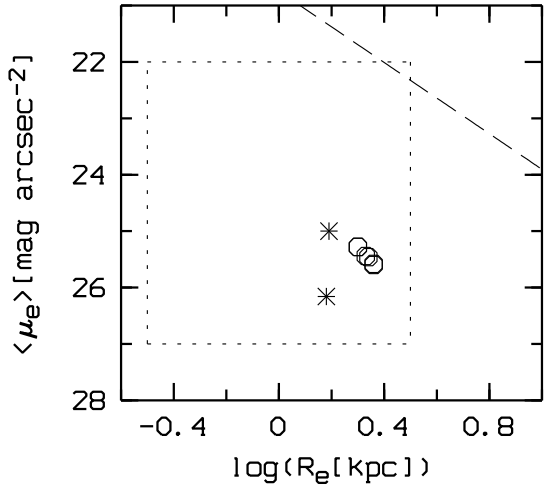,angle=-90,width=8cm,clip=}}
\caption{Same as in Figure 2 but for merging of two different mass progenitors.} 
\end{figure}

\begin{figure}
\centerline{\psfig{file=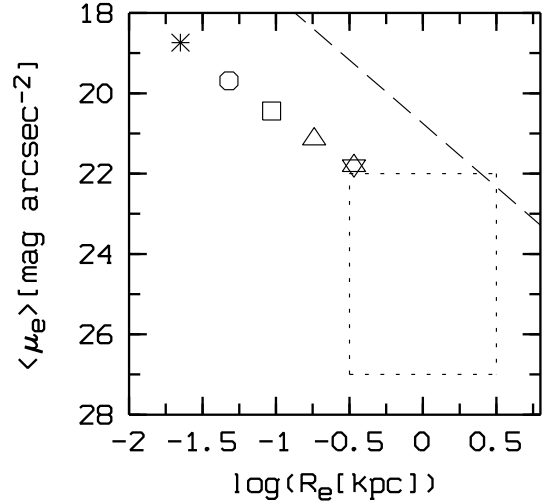,angle=-90,width=7.9cm,clip=}}
\caption{Same as in Figure 2 but for merging of ultracompact dwarf galaxies.}
\end{figure}

\begin{figure}
\centerline{\psfig{file=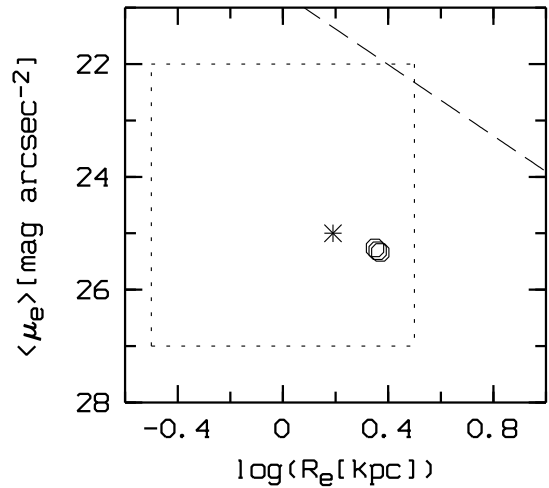,angle=-90,width=8cm,clip=}}
\caption{Same as in Figure 2 but for merging of two equal mass progenitors with dark halo.}
\end{figure}

\begin{figure}
\centerline{\psfig{file=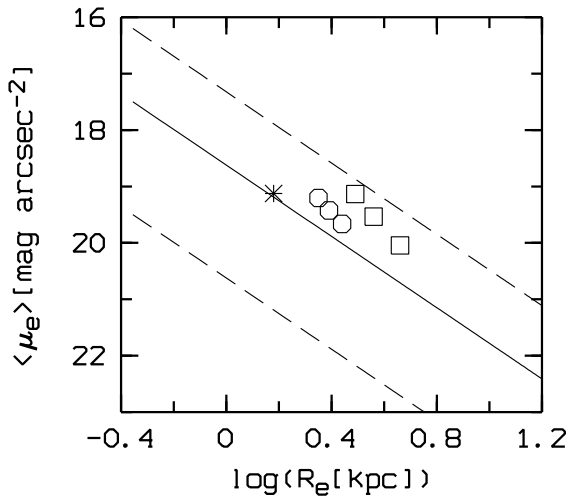,angle=-90,width=8cm,clip=}}
\caption{Merging of ellipticals lying on the KR. * - the initial progenitor.
Circles - the 1st generation progenitors; squares - the second
generation progenitors. The dashed lines correspond to those in Fig.1.}
\end{figure}

\begin{figure}
\centerline{\psfig{file=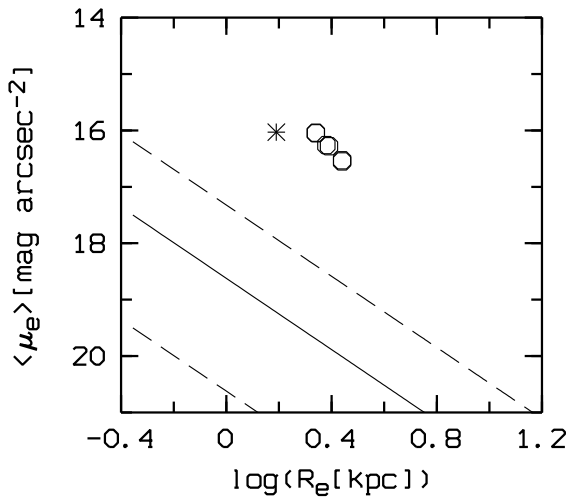,angle=-90,width=8cm,clip=}}
\caption{Merging of ellipticals lying above the KR in the
$\langle \mu_{e} \rangle - R_{e}$ plane. * - the initial progenitor.
Circles - the 1st generation progenitors. The solid line represents the KR.
The dashed lines correspond to those in Fig.1.}
\end{figure}

\section{Results}

We followed the procedure described in Capelato, de Carvalho \&
Carlberg (1995) to quantify the observables involved in the
KR: the effective radius $R_e$, containing half of the total projected
mass of the system, and the mean surface density within $R_e$,
$\Sigma_e = M(<R_e)/\pi R_e^2$, so that $\langle \mu_{e} \rangle =
-2.5 \rm{log} \Sigma_e$.  These quantities were estimated as the
median of 500 random projections.  The errors quoted for each
parameter are the rms based on the quartiles of the distribution
defined by the 500 random projections.  Columns (7) and (8) in Tables
2-7 list $R_e$ the effective radius, and $\langle \mu_{e} \rangle$,
the mean surface brightness inside $R_e$, for the merger remnant of
each simulation. These quantities are shown in Figures 2-7.

In Table 2, the values of $\hat E$ and $\hat L$ defining the
encounter span the region of rapid merging in the $\hat E - \hat L$
diagram (Binney \& Tremaine 1987). Although
arbitrary, this choice allows us to examine a particular scenario
where mergers occur over a short timescale. Some global trends can be
seen from this set of simulations. First, $\langle \mu_{e} \rangle$
always increases (lower surface brightness) with $R_e$, and the rate of
increase depends on the initial orbital energy.  Second, the presence
of angular momentum in the initial conditions does not significantly change
 the properties of the merger remnants, a trend already
noted by Capelato, de Carvalho, \& Carlberg
(1995). Therefore, we restrict ourselves to a smaller
range of $\hat L$ for each $\hat E$ value in the further simulations
described here.  We use the same $(\hat E,\hat L)$ for the simulations
presented in Tables 3-7  as for those in Table
2, except for the $\hat E$ = -1 case included in Table 2, for which
$T_{per}$ is very large compared to the other simulations.

In the simulations presented in Tables 3-7 we restrict ourselves
only to the first generation (except for the merging of objects lying
on the KR - Table 6, and the merging of UCD models - Table 4), since the 
objects produced
in the first generation tend always to move in the same direction in
the $\langle \mu_{e} \rangle - R_{e}$ plane as the first generation
systems listed in Table 2. Judging by the behavior of the the second
and third generation families displayed in Figure 2, we can safely
assume that further merging will also move the end-products to the bottom right 
in the $\langle \mu_{e} \rangle - R_{e}$ plane.

\begin{figure*}
\centerline{\psfig{file=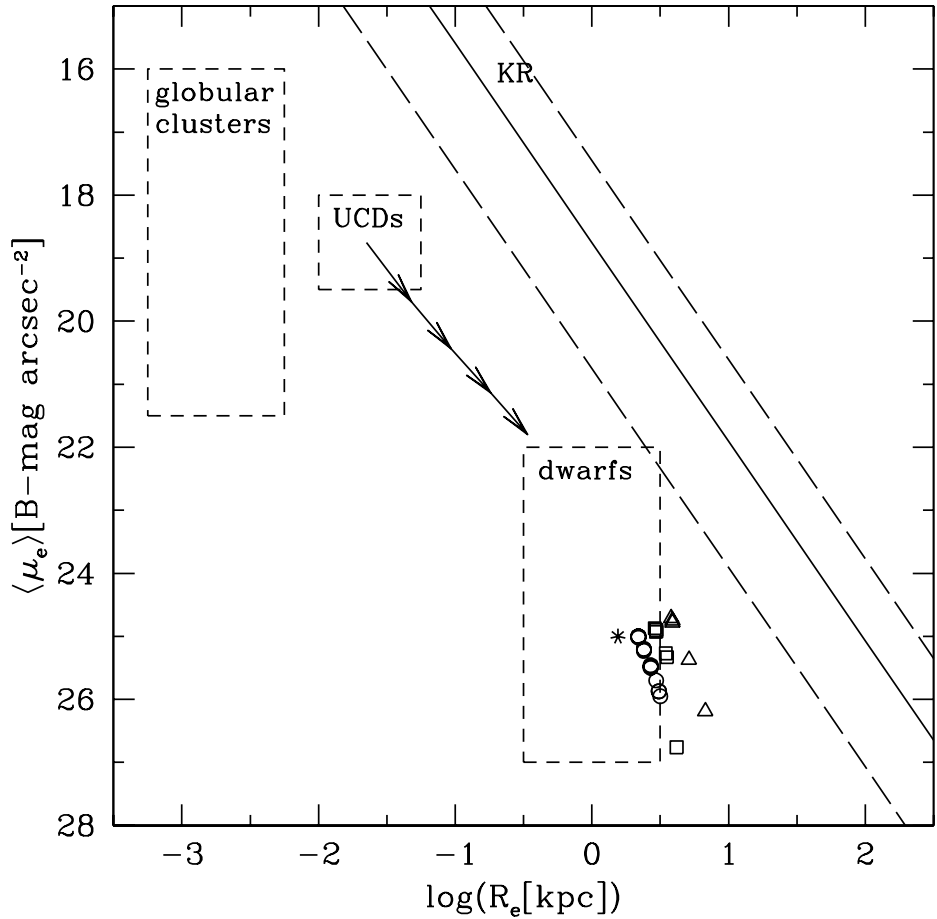,width=13cm,clip=}}
\caption{The evolution of dwarf galaxies in the $\langle \mu_{e} \rangle
- R_{e}$ plane. Merging of ultracompact dwarf galaxies is shown by large
arrows. Different symbols represent the evolution of a dwarf galaxy due to
the effects of dissipationless encounters
for three subsequent merger generations.
The solid line represents the Kormendy relation.
Both long-dashed lines correspond to those in Fig.1 which define the
lower and the upper boundaries for bright ellipticals.}
\end{figure*}

\section{Discussion}

In Capelato, de Carvalho \& Carlberg (1995), the progenitors were
systems of M=$10^{11} M_{\odot}$, while in this work we have used
low-mass progenitors ranging from 0.6$\times10^{8}$ to
9$\times10^{8}M_{\odot}$, since the main goal was to study whether or
not the low-mass systems also move along the FP. However, the loci of
the progenitors studied here are not along the KR (a projection of the
FP).  From our simulations, we see that present-day ellipticals cannot
be formed by merging present-day dwarfs, unless a considerable amount
of dissipation is involved. This result, together with other
inconsistencies present in the properties of dwarfs and bright
galaxies (see Tosi 2003), makes it nearly impossible that the major
building blocks of present-day ellipticals had properties similar to
those of dwarfs galaxies observed today.

The global trend observed in Figures 2-7 is that the end-products move
down and right in the $\langle \mu_e \rangle - R_e$ plane. On one
hand, this shows that the only way to merge dwarfs and obtain bright
ellipticals is by invoking a large amount of dissipation. On the other
hand, the same trend suggests the possibility of making dwarfs by
merging UCDs (clearly shown in Figure 4 and Figure 8), as well as the
possible formation of UCDs and dwarfs by merging globular clusters.
Fellhauer \& Kroupa (2002) have also shown that objects such as UCDs
can be formed by merging globular clusters.  Interestingly, in the
dissipationless simulations by Dantas et al. (2003), the line going
through the positions of the mergers of all of their three generations
with $-4 < \hat E < +0.5$ has the same slope as the KR.  However,
their simulations do not reconstruct the KR entirely. In essence, all
these results show how dissipationless merging moves the systems in
the $\langle \mu_e \rangle - R_e$ plane, regardless of their mass.

The structural parameters of the end-products seem to strongly depend
on the initial orbital parameters ($\hat E$, $\hat L$, and $A$). 
A recent study done by
Khochfar \& Burkert (2003) on the orbital parameters of major mergers
of CDM halos indicates that most of the encounters are nearly
parabolic. However, as shown by simulations E11, E12, and E13,
which correspond to parabolic and hyperbolic orbits, the trend is
the same as observed for the other cases, namely, the
end-products still move down and right in the $\langle \mu_e \rangle -
R_e$ plane. The global trend is independent of the $\hat E$ range,
no matter what kind of orbits we consider. 

The simulations where dark halos were added to the low-mass spherical
galaxies show that the merger remnants also shift down-right in the
$\langle \mu_e \rangle - R_e$ plane (Table 5 and Figure 5), indicating
that the presence of dark halos does not make any appreciable
difference as far as the global trend we find is concerned.

We also simulated the merging of a pair of objects lying on the KR.
The end-products also lay on the KR within the scatter of the observed
KR, and taking into account the simulation errors (Table 6 and Figure
6), are in agreement with the results obtained by Capelato, de Carvalho \&
Carlberg (1995), and Dantas et al. (2003). The simulations where the
progenitors were above the KR relation yielded the same overall
result, moving down and to the right in the $\langle \mu_e \rangle -
R_e$ plane.

In Figure 8 we show with different symbols the evolution of a progenitor
due to the effects of dissipationless encounters (see
Table 2 and Figure 2) for three
subsequent merger generations.  The resultant vectors point to the
locus of the bright ellipticals defining the KR.  From the scheme
presented in Figure 8 we see that we cannot obtain a bright elliptical
by merging low surface brightness ($\langle \mu_e \rangle = 25-26$)
dwarfs.  However, it is possible to reach the KR for bright
ellipticals by merging higher surface brightness dwarf galaxies
($\langle \mu_e \rangle = 22-23$) in a few (two or three) merger
episodes, in agreement with Bower, Kodama \& Terlevich (1998).  We can
imagine that at the initial merging stages, the dwarfs' main
interaction process is due to dissipational effects. In the $\langle
\mu_e \rangle - R_e$ diagram, this process is primarily represented by
the dwarfs' upward movement.  Then, when these dwarfs reach the upper
part of the dwarf locus, they will move according to the scheme in
Figure 8.

There are some caveats that should be noted.  Dissipation could have a
significant impact on the final position of an end-product. Also, it
is important to bear in mind that our simulations start from
equilibrium systems representing a realization of a system which was
previously formed.  Moreover, the present merger rate differs from the
one which was in the past.

We plan on doing more extensive simulations using different models for
the initial low-mass ellipticals (e.g. the exponential model or the
more general Sersic's law) as well as those for the dark halo. To
study how the initial spin of our models could affect the end-products
is also an important part of the future development.

\section*{Acknowledgements}
E.A.E. was supported by Grants for Young Scientists PD~0.2-1.2-1 and
MK - 2671.2003.02.  E.A.E. would like to thank V.P. Reshetnikov, N.Ya.
Sotnikova, and S.A. Rodionov for their assistance. RdC would like to
thank Roy Gal and Gary Mamon for their careful reading of the
manuscript, which helped improving the content of this paper. We thank
L. Hernquist for providing us with the code for realizations of the
progenitors used in our simulations. We are also grateful to
M. Drinkwater for the data on UCDs he shared with
us. H.V.C. acknowledges financial support provided by CNPq and FAPESP.
We thank the anonymous referee for very useful suggestions which helped 
to clarify some issues discussed in this paper.

\bsp

\end{document}